\documentclass[%
 reprint,
superscriptaddress,
nofootinbib,
 amsmath,amssymb,
 aps,
]{revtex4-2}
\pdfoutput=1

\usepackage{graphicx}
\usepackage[usenames,dvipsnames]{xcolor}
\usepackage{dcolumn}
\usepackage{bm}
\usepackage{bbm}
\usepackage{hyperref}
\usepackage{float}

\usepackage{physics}

\usepackage{amsthm}
\usepackage{chngcntr}
\usepackage{apptools}
\usepackage{amsmath}
\usepackage{mathtools}
\renewcommand{\Tr}[1]{\text{Tr}\left[{#1}\right]}

\usepackage{algorithm}
\usepackage{algpseudocode}
\usepackage{cleveref}

\hypersetup{colorlinks=true}
\newtheorem{myalgorithm}{Algorithm}

\newcommand{\orcidauthor}[2]{\author{{\hypersetup{hidelinks}\href{https://orcid.org/#1}{#2}}}}

 \date{\today}

\begin{document}
\title{Quantum Finite Temperature Lanczos Method}
\orcidauthor{0000-0002-5891-3289}{Gian Gentinetta}
\email{gian.gentinetta@epfl.ch}
\affiliation{Institute of Physics, \'Ecole Polytechnique F\'ed\'erale de Lausanne (EPFL), CH-1015 Lausanne, Switzerland}
\affiliation{Center for Quantum Science and Engineering, EPFL, Lausanne, Switzerland}

\orcidauthor{0000-0002-4745-7329}{Friederike Metz}
\affiliation{Institute of Physics, \'Ecole Polytechnique F\'ed\'erale de Lausanne (EPFL), CH-1015 Lausanne, Switzerland}
\affiliation{Center for Quantum Science and Engineering, EPFL, Lausanne, Switzerland}
\orcidauthor{0000-0002-2778-1703}{William Kirby}
\affiliation{IBM Quantum, IBM T.J. Watson Research Center, Yorktown Heights, NY 10598, USA}
\orcidauthor{0000-0002-8887-4356}{Giuseppe Carleo}

\affiliation{Institute of Physics, \'Ecole Polytechnique F\'ed\'erale de Lausanne (EPFL), CH-1015 Lausanne, Switzerland}
\affiliation{Center for Quantum Science and Engineering, EPFL, Lausanne, Switzerland}

\begin{abstract}
The computation of thermal properties of quantum many-body systems is a central challenge in our understanding of quantum mechanics. We introduce the Quantum Finite Temperature Lanczos Method (QFTLM), which extends the finite-temperature Lanczos method to quantum computers by combining real-time quantum Krylov methods with efficient preparation of typical states for trace estimation. This approach enables the computation of thermal expectation values while avoiding the exponential scaling inherent to classical exact simulation techniques. Numerical experiments on the transverse-field Ising model show that QFTLM can reproduce thermal observables over a wide temperature range. We further analyze the influence of Krylov dimension, number of trace-estimator states, and Trotter error, and show that suitable regularization is essential for robustness in noisy settings. These results establish QFTLM as a promising framework for finite-temperature quantum simulation.
\end{abstract}

\maketitle
\section{Introduction}

Understanding properties of quantum many-body systems at finite temperature is fundamental in advancing our knowledge of condensed matter physics and quantum chemistry~\cite{mermin_thermal_1965, georges_dynamical_1996, white_finite-temperature_2020}. To effectively describe experimental phenomena, thermal observables such as magnetization, specific heat and correlation functions are essential. However, performing such calculations remains a central challenge in modern computational quantum physics due to the Hilbert space dimension growing exponentially with the system size. The finite-temperature regime is especially challenging because observables depend on the full spectrum of the Hamiltonian, rendering many computational tricks employed in ground state search ineffective in this setting~\cite{harsha_thermofield_2019, sun_finite-temperature_2020}. 

Nevertheless, classical simulation techniques such as Quantum Monte Carlo methods and tensor networks have made significant progress towards accessing finite-temperature properties~\cite{sinha_efficient_2024}. In particular, the Finite Temperature Lanczos Method (FTLM) has emerged as a promising candidate~\cite{jaklic_lanczos_1994}. FTLM combines ideas of trace estimation~\cite{golub2010matrices,avron2011randomized,golub2013matrix,Roosta-Khorasani2015improved,ubaru2017fast,martinsson2020randomized, hutchinson_stochastic_1989}, where distributions of \emph{typical states} are used to estimate thermal expectation values, with Krylov methods that perform exact diagonalization within a relevant subspace of the exponentially large Hilbert space. While the diagonalization happens in the Krylov subspace, FTLM still requires the preparation of the Krylov basis states in the full Hilbert space. To scale to larger system sizes, classical FTLM thus has to be combined with additional simulation methods, such as tensor networks, that allow for an efficient representation of many-body quantum states~\cite{sinha_efficient_2024}.

Quantum computers offer a natural framework for the simulation of many-body quantum systems as the number of qubits required to capture the Hilbert space scales polynomially in the number of particles, avoiding the exponential scaling encountered in exact classical simulation methods~\cite{Feynman1982,ibm_17, chiesa_2019, google_2020, arute2020observation, neill2021}. Furthermore, there has been remarkable progress in using quantum computers to construct Krylov subspaces for the computation of ground and excited state problems~\cite{mcclean2017subspace,colless2018computation,parrish2019filterdiagonalization,motta2020qite_qlanczos,takeshita2020subspace,huggins2020nonorthogonal,stair2020krylov,urbanek2020chemistry,kyriienko2020quantum,cohn2021filterdiagonalization,yoshioka2021virtualsubspace,epperly2021subspacediagonalization,seki2021powermethod,bespalova2021hamiltonian,baker2021lanczos,baker2021block,cortes2022krylov,klymko2022realtime,jamet2022greens,baek2023nonorthogonal,tkachenko2022davidson,lee2023sampling,zhang2023measurementefficient,kirby2023exactefficient,shen2023realtimekrylov,yang2023dualgse,yang2023shadow,ohkura2023leveraging,motta2023subspace,anderson2024solving,kirby2024analysis,yoshioka2024diagonalization,byrne2024super,yu2025quantum, piccinelli_quantum_2026}. A recent work connecting quantum Krylov methods to Szegö quadrature rules describes how thermal observables can be expressed in a Krylov subspace generated through real-time dynamics~\cite{kirby_quantum_2025}.

In this work, we combine the framework presented in~\cite{kirby_quantum_2025} with the efficient preparation of trace estimator states to define the Quantum Finite Temperature Lanczos Method (QFTLM) algorithm. We show that this method allows us to compute thermal properties over a wide temperature range. In numerical experiments, we analyze how the algorithm scales with system size, Krylov dimension, number of trace estimator states as well as how Trotterization of the dynamics affects the final result.

The remainder of this paper is organized as follows. In~\Cref{sec:methods}, we review the classical FTLM and discuss how we extend it to the quantum case. We then test our method on the transverse-field Ising model in~\Cref{sec:results} where we also investigate the effect of noise on our results.

\section{Methods}
\label{sec:methods}
\subsection{Finite Temperature Lanczos Method}
The Lanczos method is a widely used technique to obtain the ground state and low-lying excited states of sparse Hamiltonians~\cite{lanczos1950iteration}. Extending this algorithm to compute thermal properties of quantum systems is non-trivial, as those generally depend on the full spectrum of the Hamiltonian, which cannot be captured effectively in a single Krylov subspace. The Finite Temperature Lanczos Method (FTLM) addresses this challenge by exploiting properties of \textit{typical states}~\cite{jaklic_lanczos_1994, iitaka_random_2004, hutchinson_stochastic_1989}, known in the classical numerical linear algebra literature as \textit{trace estimators}. Typical states $\ket{\psi_k}$ can be used to estimate expectation values of thermal properties, i.e.~\cite{schnack_accuracy_2020}
\begin{equation}
\label{eq:ftlm}
    \frac{\Tr{e^{-\beta H} O}}{\Tr{e^{-\beta H} }} \approx  \frac{\sum_{k=1}^K\bra{\psi_k} e^{-\beta H} O \ket{\psi_k}}{\sum_{k=1}^K\bra{\psi_k} e^{-\beta H} \ket{\psi_k}}.
\end{equation}
In fact, there is numerical evidence that for large systems with dense spectra at high temperatures, such observables can be accurately estimated using a single random state~\cite{schnack_magnetism_2018}. Building on this, the FTLM constructs $K$ distinct Krylov subspaces starting from typical states $\ket{\psi_k}$ in which the projected Hamiltonian is then diagonalized to obtain the projected thermal state.

\subsection{Quantum Finite Temperature Lanczos Method}

In this subsection we discuss how quantum computers can be employed to perform the FTLM. We discuss how to efficiently prepare typical states and evolve them in time to generate a real-time Krylov basis. We then describe the measurements required to project the thermal state and observables into the subspace.

Ideally, we would like to prepare states drawn randomly from the Haar distribution over the unitary group on $n$ qubits~\cite{shen_simple_2025}. However, preparing truly random states on a quantum device requires circuit depth exponential in the number of qubits~\cite{schuster_random_2025, aaronson_complexity_2016}. Fortunately, there are several approaches to efficiently approximating the Haar distribution up to a moment $k$, such that for any quantum state $\ket{\psi}$
\begin{equation}
    \mathbb{E}_{U \sim p_k}\left[\left(U\ket{\psi}\bra{\psi}U^\dagger\right)^{\otimes k}\right] = \mathbb{E}_{U \sim\text{Haar}}\left[\left(U\ket{\psi}\bra{\psi}U^\dagger\right)^{\otimes k}\right],
\end{equation}
where the distribution $p_k$ is usually referred to as a \textit{k-design}~\cite{hayden_randomizing_2004,dankert_exact_2009, gross_evenly_2007, harrow_quantum_2009}. A particularly efficient approach is described in~\cite{schuster_random_2025}, where random Clifford circuits are patched together to implement a $k$-design in $O(\log(n))$ depth.

Instead of preparing fully random states, it is shown in~\cite{shen_simple_2025} that the trace estimator converges also for ($k$-designs of) random diagonal states
\begin{equation}
\ket{\psi_{\text{Q-Hutch}}} = \frac{1}{\sqrt{2^n}}\sum_{k=1}^{2^n}e^{-i\phi_k}\ket{k}
\end{equation}
with random complex phases $\phi_k \sim \text{Uniform}(0,2\pi)$. These states are known as \textit{Quantum Hutchinson} states~\cite{shen_simple_2025} and extend the classical Hutchinson states, where the random phases are real~\cite{hutchinson_stochastic_1989}. A 3-design of this distribution can be implemented in $O(n)$ depth using time-evolution circuits of spin glass Hamiltonians~\cite{shen_simple_2025}. The combination of this simple circuit construction and convergence guarantees stating that the trace estimator scales with $O(1/\sqrt{K})$ motivates our use of these estimators.

In the next step, we construct a Krylov subspace for each $\ket{\psi_k}$ in which we then compute the expectation values in the numerator and denominator of~\Cref{eq:ftlm}. In classical Krylov methods, the subspace is typically created by repeatedly applying $H$ to the initial quantum state. While this operation is possible in principle on a quantum computer, it requires fault-tolerant techniques such as qubitization~\cite{Low2019}. Instead, we use the quantum Krylov algorithm for Szegö quadrature described in~\cite{kirby_quantum_2025}. The algorithm consists of two main steps. First, a real-time Krylov basis is constructed by evolving $\ket{\psi_k^0} := \ket{\psi_k}$ forward and backward in time using Trotterization of $e^{-i\Delta t H}$. We denote the $D=2d + 1$ non-orthogonal basis states as
\begin{equation}
|\psi_k^j\rangle = \exp(-i\Delta tH)^j\ket{\psi_k^0}, \quad    j \in \{-d,-d+1,\dots , d\},
\end{equation}
where the time step $\Delta t$ is chosen such that $E_j \Delta t \in [-\pi, \pi)$ for all energy eigenvalues $E_j$, i.e. $\Delta t \leq \pi/||H||$.

We then measure the complex Gram matrix ${S_{ij} = \bra{\psi^i}\ket{\psi^j}}$ (from here on we drop the subscript $k$ on $\ket{\psi_k^i}$ for readability) using the Hadamard test~\cite{Cleve_1998, aharonov2006jonespolynomial}. Noting that $U_{ij} = \bra{\psi^i}\exp(-i\Delta tH)\ket{\psi^j} = S_{ij+1}$ we can directly obtain the projected time-evolution unitary. The Gram matrix and the projected unitary are regularized in post-processing and the Krylov basis is orthonormalized using $S$ (see~\Cref{sec:regularization}) to obtain the unitary $\mathbf{\tilde U}$ in the orthonormalized Krylov subspace. We can then diagonalize $\mathbf{\tilde U}$ to obtain the eigenvectors $V$ and eigenvalues $\Lambda = \text{diag}(\lambda_1, \dots \lambda_D)$ of the time-evolution unitary in the Krylov subspace. Noting that $\lambda_j = e^{-iE_j\Delta t}$, we can directly read out the energy eigenvalues $E_j$ and reconstruct the thermal state in the Krylov subspace. In matrix form, the expectation value is then given as
\begin{equation}
    \bra{\psi^0}e^{-\beta H}\ket{\psi^0} \approx \left[S^{1/2} V \Lambda^{-i\beta/\Delta t} V^\dagger S^{1/2}\right]_{0,0}.
\end{equation}
A detailed description of the algorithm is provided in~\cite{kirby_quantum_2025}.

To estimate $\bra{\psi^0}e^{-\beta H}\hat O\ket{\psi^0}$ for an observable $\hat O$ that is not diagonal in the energy eigenbasis, we also have to measure $\bra{\psi^j}\hat O\ket{\psi^0}$ on the quantum device to project the observable into the Krylov subspace. Noting that the projector into the Krylov basis can be written as $\Pi = \sum_{i,j}S^{-1}_{ij}\ket{\psi^i}\bra{\psi^j}$, we find
\begin{align}
    \bra{\psi^0}e^{-\beta H}\hat O\ket{\psi^0} &\approx \bra{\psi^0}e^{-\beta H}\Pi\hat O\ket{\psi^0} \\
    &= \left[S^{1/2}f(\mathbf{\tilde U})S^{-1/2} O\right]_{0,0}.
\end{align}

Among quantum Krylov methods, this algorithm is particularly useful in our case, as it admits convergence guarantees for approximating Gibbs states. In particular, it is shown in~\cite{kirby_quantum_2025} that the error in approximating $\bra{\psi_k}e^{-\beta H}\ket{\psi_k}$ is exponentially suppressed with the subspace dimension.

\begin{figure}
    \centering
    \includegraphics[width=0.8\linewidth]{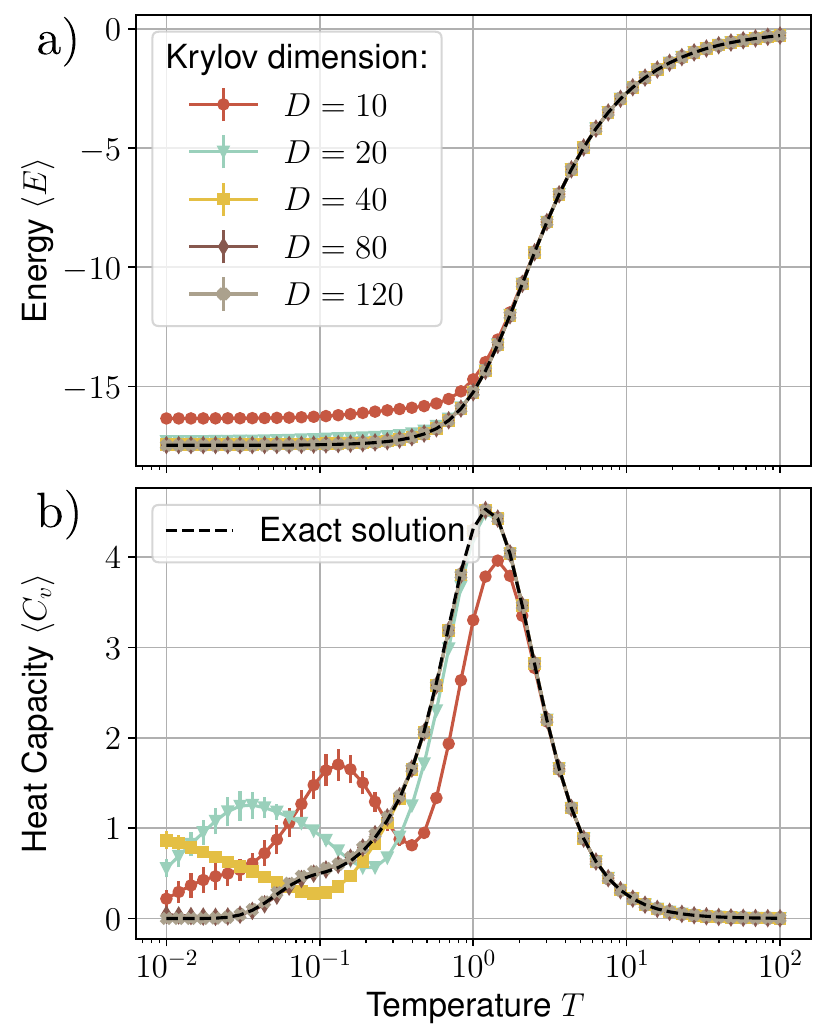}
    \caption{Thermal expectation values of the energy (a) and heat capacity (b) as a function of the temperature for $L=14$ spins in the TFIM (see~\Cref{eq:tfim}). We compare the results obtained with QFTLM for varying Krylov dimensions $D$ with the exact solution (in black). The real-time Krylov basis is approximated with 4 first-order Trotter steps and the expectation values are computed by averaging over $K=40$ quantum Hutchinson states. In the low-temperature limit, choosing a smaller Krylov subspace leads to significant deviations from the exact solutions. For high temperatures, all simulations qualitatively align with the ground truth.}
    \label{fig:energy_and_heat}
\end{figure}

\begin{figure*}
    \centering
    \includegraphics[width=0.8\linewidth]{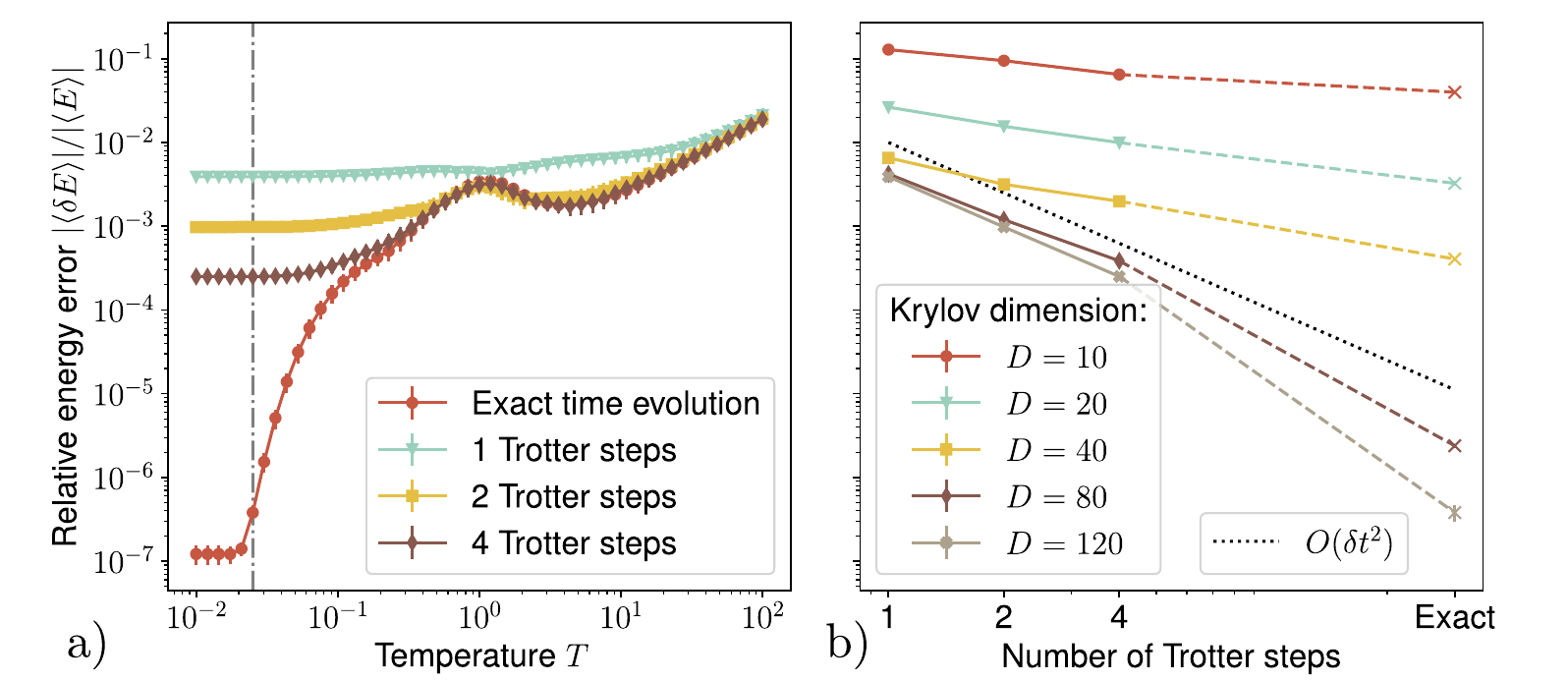}
    \caption{Propagation of Trotter error through the QFTLM protocol. In panel (a), we plot the relative error of the thermal energies as a function of temperature for QFTLM simulations with varying number of Trotter steps used to build the real-time Krylov basis. The Krylov dimension is fixed to $D=120$ and we average over $K=40$ quantum Hutchinson states. At low temperatures, the Trotter error is clearly noticeable whereas at high temperatures other sources dominate the overall error on the thermal energy. In panel (b), we explicitly plot the scaling of the relative energy error as a function of the number of Trotter steps in the low-temperature limit ($T=0.025$, indicated by the gray dashed line in the left plot). For a large enough Krylov dimension $D$, the error scales roughly as $O(\delta t^2)$ (indicated by the black dotted line), where $\delta t$ denotes the Trotter time step.} 
    \label{fig:trotter_scaling}
\end{figure*}

\subsection{Regularization}
\label{sec:regularization}

In practice, the Gram matrix $S$ is often ill-conditioned, leading to numerical instabilities. This becomes an even greater problem if the entries $S_{ij}$ and $U_{ij}$ are measured under noisy conditions. Even under perfect conditions, the matrix $U$ constructed by measuring $U_{ij} =\bra{\psi^i}\exp(-i\Delta tH)\ket{\psi^j}$ is not a priori unitary. For those reasons, we must regularize the measured matrices and orthonormalize the Krylov basis before proceeding with the QFTLM algorithm. A regularization scheme was proposed in~\cite{kirby_quantum_2025}, but we find that in practice, it is useful to include an additional step in the regularization, which we now describe. Pseudocode is given in~\Cref{algo:regularization}.
 
As described in~\cite{kirby_quantum_2025}, we start by orthonormalizing the Krylov subspace. First, the Gram matrix is regularized with a Tikhonov shift $\eta$, ensuring its eigenvalues are at least of size $\eta$. In practice, it is advised to choose $\eta$ in the order of magnitude of the expected noise on the measured overlaps. We can then safely compute the inverse square root of $S$ and obtain the representation of $U$ in an orthonormal basis for the Krylov space~\cite{lowdin_nonorthogonality_1970}:
\begin{equation}
    \mathbf{\tilde U'} = S^{-1/2}US^{-1/2}.
\end{equation}

At this point, our regularization procedure deviates from the analysis in~\cite{kirby_quantum_2025}. As discussed in~\cite{kirby_quantum_2025}, in the noiseless case all but one of the eigenvalues of the resulting matrix $\mathbf{\tilde U'}$ should now have unit length. In practice however, we observed that noisy inputs often lead to multiple smaller eigenvalues. In~\cite{kirby_quantum_2025}, $\mathbf{\tilde U'}$ is simply projected to the nearest unitary operator (which is unique), but we find that an additional step improves performance in practice.
Prior to projecting to the nearest unitary, we prune singular values of $\mathbf{\tilde U'}$ that lie below a threshold parameter $\mu$. This is achieved by performing a singular value decomposition $\mathbf{\tilde U'} = P\Sigma Q^\dagger$ and transforming $\mathbf{\tilde U'}$ with the truncated (and hence non-square) transformation $P$
\begin{equation}
\label{eq:svd_trunc}
    \mathbf{\tilde U''}_{ij} = \sum_{\substack{k,l\\ \text{s.t.}\ \sigma_k,\sigma_l\geq\mu}}P^\dagger_{ik}\mathbf{\tilde U'}_{kl}P_{lj}.
\end{equation}
The motivation for this is that with the exception of the one singular value that is ``supposed'' to deviate below 1, the deviations of additional singular values indicate that the corresponding dimensions are corrupted by noise, with the degree of the deviation serving as a proxy the degree of the corruption.
In our numerics, the optimal threshold $\mu$ is typically found to be close to 1.

After this step, we can proceed with the algorithm outlined in~\cite{kirby_quantum_2025} by projecting $\mathbf{\tilde U''}$ onto the nearest unitary. To this end, we perform another singular value decomposition $\mathbf{\tilde U''} = WDR^\dagger$ and set all singular values to unity, yielding $\mathbf{\tilde U} = WR^\dagger$. Finally, we note that the eigenvectors obtained through diagonalizing $\mathbf{\tilde U} = V'\Lambda V$ have to be transformed back into the original basis $V = PV'$. 

We note that the optimal regularization parameters $\eta$ and $\mu$ vary significantly depending on the input noise. In the results section we analyze this in more detail and discuss approaches to choose these hyper-parameters effectively.

\begin{myalgorithm}[``Regularized Diagonalization"]~
\label{algo:regularization}
\normalfont
\begin{algorithmic}[1]

    \State\textbf{input} noisy quantum Krylov matrices $(U', S')$
    
    \State Choose $\eta>0$\Comment{Tikhonov regularization}
    \State Choose $\mu>0$\Comment{Singular value truncation}

    \State $S =  S'+\eta I$\label{line:tikhonov}

    \State $\mathbf{\tilde U'} \gets S^{-1/2}U'S^{-1/2}$ \Comment{Orthonormalization}

    \State $P'\Sigma Q^\dagger=\mathbf{\tilde U'}$ \Comment{Singular value decomposition}
    \State $P \gets P'[:, \Sigma > \mu]$\Comment{Singular value truncation}
    \State $\mathbf{\tilde U''} \gets P^\dagger \mathbf{\tilde U'}P$\Comment{Transform into truncated basis}
    \State $WDR^\dagger=\mathbf{\tilde U''}$ \Comment{Singular value decomposition}
    \State $\mathbf{\tilde U} \gets WR^\dagger$ \Comment{Project to nearest unitary}
    \State $\Lambda = \text{eigvals}(\mathbf{\tilde U})$\Comment{Diagonalize}
    \State $V' = \text{eigvecs}(\mathbf{\tilde U})$
    \State $V \gets PV'$\Comment{Transform back into the full basis}

    \State\textbf{return} $\Lambda,V$
    
\end{algorithmic}
\end{myalgorithm}

\section{Results}
\label{sec:results}

We use classical simulations to study the performance of the QFTLM on the transverse-field Ising model (TFIM) given by the Hamiltonian
\begin{equation}
\label{eq:tfim}
    H = \sum_{i=1}^L X_i - \sum_{i=1}^{L-1}Z_i Z_{i+1},
\end{equation}
where $X_i$, $Z_i$ are the Pauli operators acting on qubit $i$. In this section, we investigate how the different sources of error (finite Krylov dimension, trace estimation, Trotter approximation, and shot noise) influence the accuracy of the simulation and how the error scales with the system size $L$.

In~\Cref{fig:energy_and_heat}, we show the thermal expectation values of the energy  $\langle E \rangle = \Tr{He^{-\beta H}}/\Tr{e^{-\beta H}}$ (panel (a)) and the heat capacity $C_v = \beta^2(\Tr{H^2e^{-\beta H}}/\Tr{e^{-\beta H}} - \langle E \rangle^2)$ (panel (b)) as functions of the temperature for $L=14$ spins and varying Krylov dimensions. To build the real-time Krylov basis, we evolve a quantum Hutchinson state in time by approximating $\exp(-i\Delta tH)$ with 4 first-order Trotter steps~\cite{hatano_finding_2005, lloyd1996quantumsimulators, childs_theory_2021} and repeating this circuit $j$ times to approximate $j$ timesteps $\exp(-i\Delta tH)^j$. 
We iterate over $K=40$ trace estimator states to compute the thermal averages. At low temperatures ($T < 1$), the results deviate qualitatively from the exact solution for smaller Krylov dimensions whereas at high temperatures all simulations overlap with the exact solution.
This supports the intuition that at high temperatures, the features in the Gibbs state are smoothed and therefore become easier to approximate at low Krylov dimension.

In~\Cref{fig:trotter_scaling}, we further analyze the effect of Trotterization in the QFTLM protocol by showing the relative error of the thermal energy expectation values as a function of the temperature for 1, 2 and 4 first-order Trotter steps used to approximate $\exp(-i\Delta tH)$. For now, the Krylov dimension is fixed at $D=120$ and we again average over $K=40$ quantum Hutchinson states. As with the Krylov dimension, the error due to choosing large Trotter steps is most apparent in the low-temperature regime. To demonstrate that further, we consider a slice at low temperature ($T = 0.025$, indicated by the dash-dotted line in panel (a)) and separately display the error as a function of the number of Trotter steps $N_T$ in panel (b). Here, we again include simulations for varying Krylov dimensions and note that for sufficiently large $D$, the error in the thermal energy roughly scales as $O(1/N_T^2)$ which corresponds to the expected error for a first-order Trotter approximation. 

\begin{figure} 
    \centering
    \includegraphics[width=0.8\linewidth]{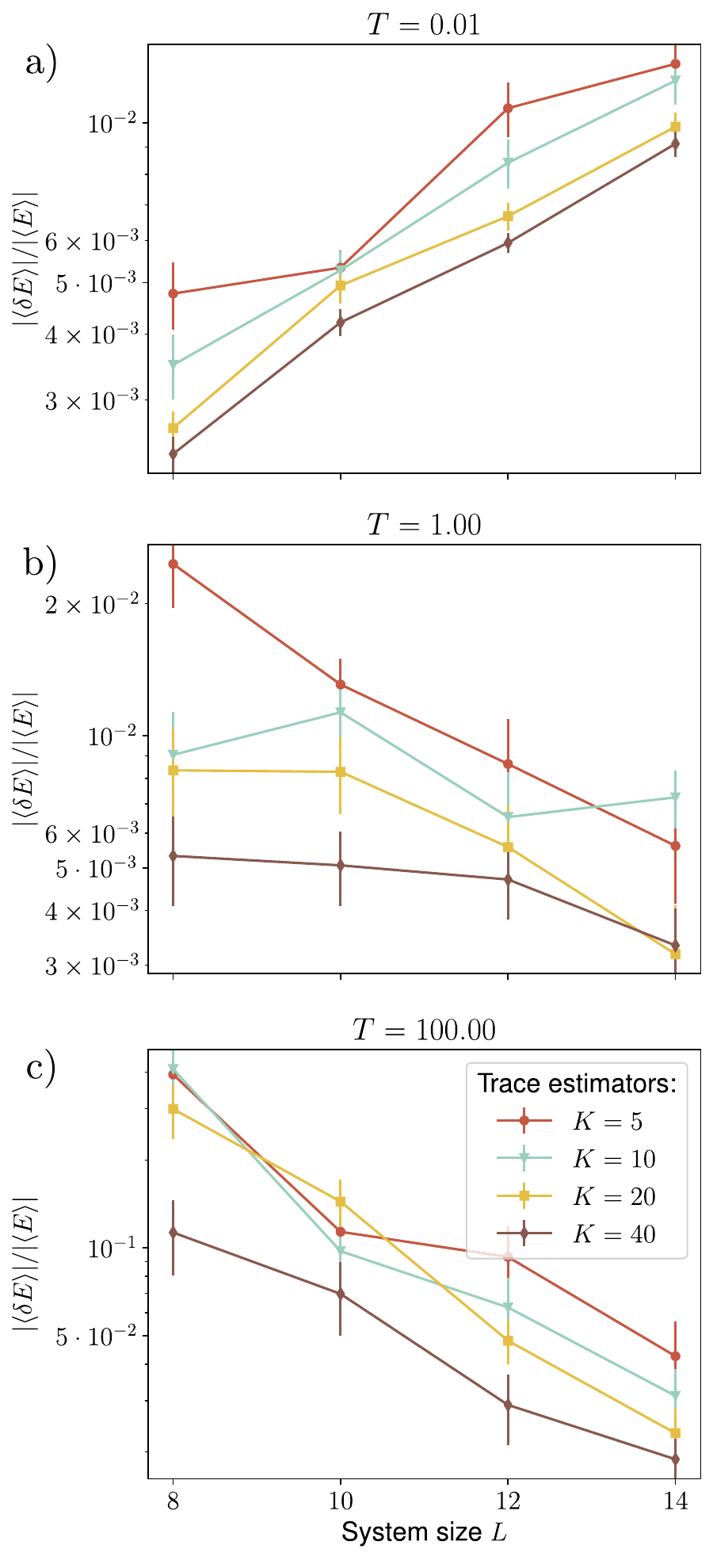}
    \caption{Scaling of the QFTLM error with system size $L$. We plot the relative error in the energy expectation values at three temperatures, $T=0.01, 1,$ and $100$, in panels (a), (b), and (c), respectively, as a function of the number of spins $L$, for a varying number of quantum Hutchinson states $K$. The Krylov subspace dimension is fixed to $D=20$ and the number of Trotter steps to 4. At low temperatures, the error at fixed Krylov dimension increases with the system size. Conversely, at high temperatures, we observe a decrease in the error as the number of spins increases.}
    \label{fig:system_size_scaling}
\end{figure}

Finally, we investigate how the performance of our method scales with the system size $L$ in the different temperature regimes. \Cref{fig:system_size_scaling} shows the relative error of the thermal energy expectation as a function of $L$ for a varying number of quantum Hutchinson states $K$ used to estimate the trace. The Krylov dimension is kept constant at $D=20$ over all experiments so that the error due to the finite Krylov dimension is noticeable even for small $L$ and we can study how it scales with the system size. In the low-temperature regime ($T=0.01$, panel (a)) we observe that the error increases with $L$. This is to be expected as for $T\to 0$ the problem essentially reduces to finding the ground state, where the error depends on the gap between the ground-state energy and the first excited state as well as on the overlap between the initial state and the ground state~\cite{epperly2021subspacediagonalization, kirby2024analysis}. Both of those properties decrease with the system size, and for typical states, which we employ as the initial states in our work, the overlap with the ground state decays exponentially. However, this effect diminishes as the temperature increases and at $T=100$ (panel (c)) the error in fact decreases with $L$. This provides evidence that, like FTLM~\cite{schnack_accuracy_2020}, QFTLM works particularly well at high temperatures and large system sizes.

\begin{figure*} 
    \centering
    \includegraphics[width=\linewidth]{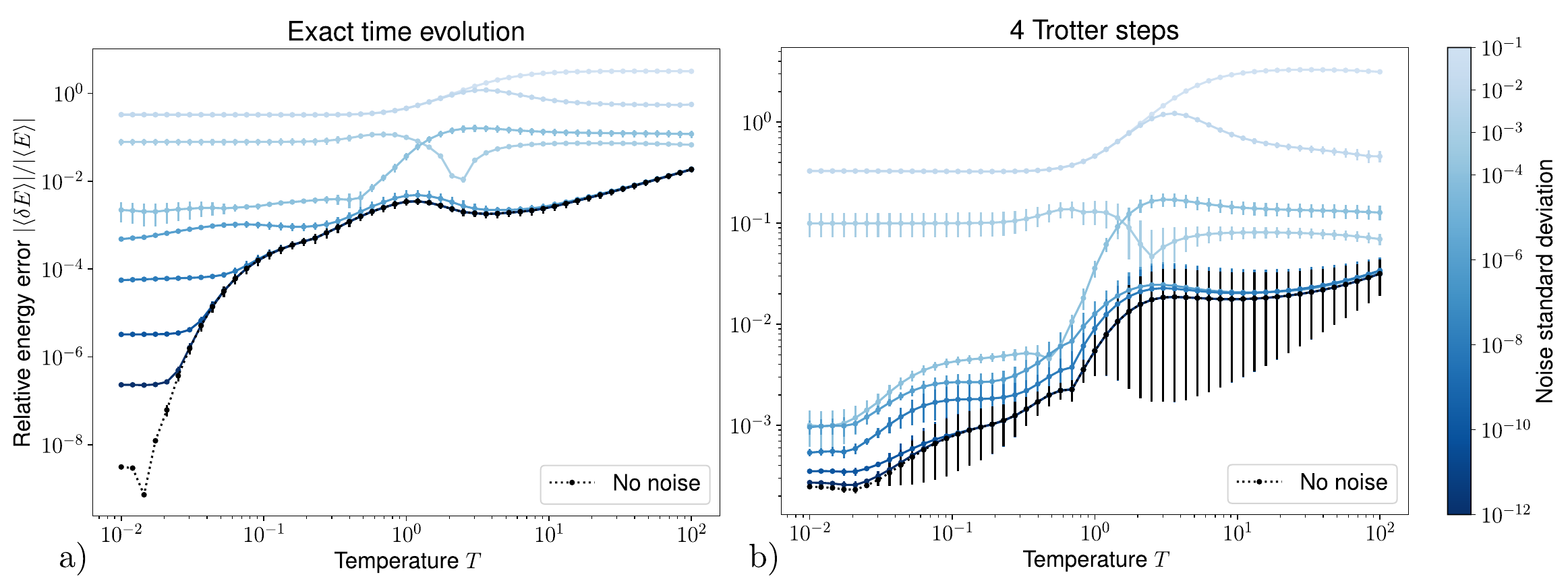}
    \caption{Effect of noise on the error of thermal energies obtained through QFTLM. Gaussian noise with varying standard deviation is added to the measured Gram matrix $S$ and projected unitary $U$. In panel (a), the time evolution is performed exactly without Trotterization. In this case, the noise is the dominant source of error at low temperatures whereas for sufficiently small standard deviations the error saturates at high temperatures. In panel (b), we perform the time evolution by splitting $e^{-iH\Delta t}$ into 4 Trotter steps. Now, the Trotter error dominates at low temperatures for small enough noise levels. Colorbar ticks indicate the simulated noise levels.}
    \label{fig:noise}
\end{figure*}

In practice, the overlaps computed on the quantum device can only be estimated to finite precision. Even on a fault-tolerant quantum computer, those quantities will be affected by finite sampling noise. To understand how the QFTLM algorithm performs under noisy conditions, we add Gaussian noise to the Gram matrix $S$ and projected unitary $U$ and observe how this error propagates through the algorithm. We again look at $L=14$ spins in the TFIM described in~\Cref{eq:tfim} for a Krylov dimension of $D=120$ and average over $K=40$ quantum Hutchinson states. In~\Cref{fig:noise}, we show the relative error of the thermal energy expectation values as a function of the temperature for varying noise levels. We show results for a simulation where the time evolution is performed exactly (panel (a)) as well as one where we approximate the $e^{-iH\Delta t}$ using 4 Trotter steps (panel (b)).
In the exact case, we observe that the noise is the dominant source of error at low temperatures.
At high temperatures ($T>1$), Gaussian noise with standard deviation of $10^{-6}$ is small enough that overall it is not the dominant source of error.
In the simulation using Trotterized time evolution, the error for noise levels below $10^{-6}$ is comparable throughout the whole temperature range since the Trotterization is the dominant source of error in this case. Still, we note that for standard deviations above $\sigma =10^{-5}$, the simulation deviates significantly from the exact solution. On an ideal quantum computer, we expect to require $O(1/\sigma^2)$ shots in order to measure the overlaps with precision $\sigma$, which implies a significant sampling overhead for the QFTLM algorithm.

\begin{figure}
    \centering
    \includegraphics[width=\linewidth]{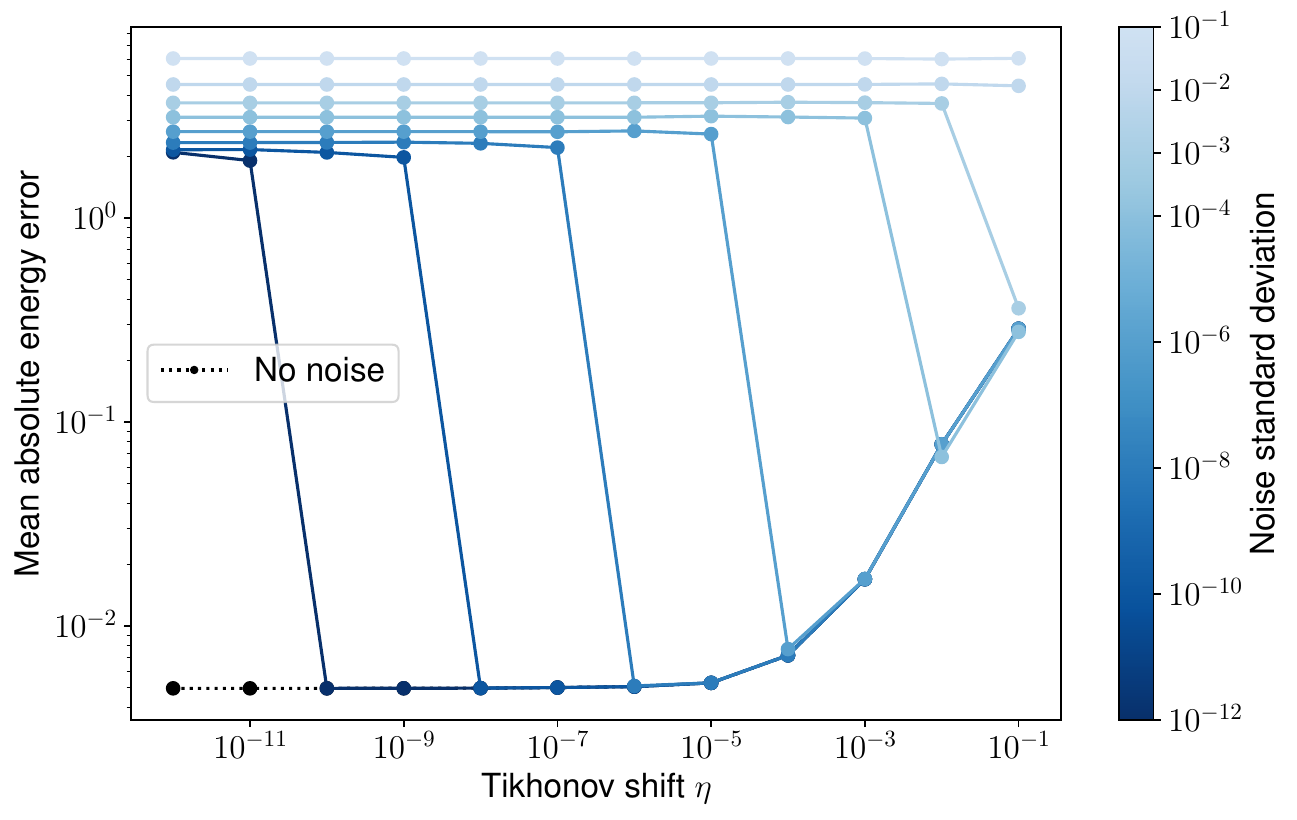}
    \caption{Importance of appropriate regularization for noisy simulations. We plot the absolute energy error averaged over all temperatures (between $10^{-2}$ and $10^2$) as a function of the Tikhonov shift $\eta$ chosen to regularize the Gram matrix (Line 4 in~\cref{algo:regularization}). We find that for each noise level there is a threshold (roughly a factor of 100 above the noise level) defining the minimal $\eta$ above which we obtain reasonable errors in the energy. Below that threshold, the error spikes up to $O(1)$, whereas above the threshold the error remains low up to $\eta \gtrsim 10^{-4}$. Colorbar ticks indicate the simulated noise levels.}
    \label{fig:Tikhonov}
\end{figure}

For simulations affected by shot noise, the regularization procedure discussed in~\Cref{sec:regularization} becomes particularly important. While throughout our experiments, we observed that choosing the singular-value truncation threshold in~\Cref{eq:svd_trunc} around $\mu \approx 0.9$ yields the best results, the ideal choice of the Tikhonov shift $\eta$ strongly depends on the noise level. In~\Cref{fig:Tikhonov} we show the mean error on the thermal energies as a function of $\eta$. We observe that the optimal result is achieved when $\eta$ is chosen two orders of magnitude larger than the standard deviation of the Gaussian noise applied to the measured overlaps. If the shift is chosen too small, the computation of $S^{-1/2}$ fails, yielding huge errors in the observed energies. On the other hand, we note that it is acceptable to choose $\eta$ to be larger than the optimal value as the error does not significantly increase until $\eta \gtrsim 10^{-4}$. Ideally, $\eta$ should thus be chosen generously above 100 times the expected noise level. It is important to note that the choice of regularization happens in post-processing and hence can be optimized given the noisy data. In particular, in cases where the ground-state energy is known (or can be computed using other computational methods), $\eta$ and $\mu$ can be fine-tuned to minimize the error on the ground state obtained in the diagonalization of $\mathbf{\tilde{U}}$. To this end, we note that obtaining the ground-state energy of a system is often well understood compared to the computation of thermal properties.

\section{Conclusion}
We introduce the Quantum Finite Temperature Lanczos Method (QFTLM), combining the classical FTLM framework with real-time quantum Krylov methods and efficient trace estimation using typical states. By averaging expectation values over multiple Krylov iterations starting from independent quantum states, we are able to compute thermal properties, effectively extending the Krylov framework beyond low-lying excited states. By employing a quantum computer to prepare the Krylov basis and measure the Gram matrix, we circumvent the curse of dimensionality encountered in classical simulation techniques.

We demonstrate through numerical experiments on the transverse-field Ising model that QFTLM can accurately reproduce thermal expectation values over a wide temperature range. Our results highlight the complementary roles of the Krylov dimension, the number of trace estimator states, and the accuracy of the time-evolution operator. We observe that the magnitude of the Trotterization error and Krylov dimension particularly matter in the low-temperature regime, whereas at high temperatures typicality becomes increasingly important. Of particular interest is the scaling of the error with the system size, where we find that at high temperatures the relative error at fixed Krylov dimension decreases as the number of spins increases. This suggests a beneficial scaling for the QFTLM algorithm once it is extended past the size of exact diagonalization.

Furthermore, we investigate how noise in the quantum measurement propagates through the QFTLM algorithm. To this end, we design a regularization procedure to separate signal from noise. We observe that the QFTLM is robust to small noise levels, but it is clear that a large number of shots will be required to resolve the measured overlaps to a precision that allows for accurate results.

In summary, QFTLM provides a framework for computing finite-temperature properties of quantum many-body systems on quantum computers. The method is applicable to arbitrary quantum Hamiltonians beyond the TFIM studied in this paper. We expect QFTLM to serve as a useful tool in estimating thermal expectations in condensed-matter physics and quantum chemistry on near- to mid-term quantum devices.

\begin{acknowledgments}
This research was supported by the NCCR MARVEL, a National Centre of Competence in Research, funded by the Swiss National Science Foundation (grant number 205602).
 \end{acknowledgments}

\section*{Code availability}
The code and data required to reproduce the figures shown in this work are available at~\cite{gentinettagianquantumfinitetemperaturelanczos_nodate}.

\bibliography{notes}
\end{document}